\documentclass[aps,prl,twocolumn,showpacs]{revtex4}
\usepackage{graphicx}
\usepackage{amssymb}
\usepackage{amsmath}

\begin{document}

\title{Creep of a fracture line in paper peeling}

\author{J. Koivisto, J. Rosti, and M.J. Alava}

\affiliation{Helsinki University of Technology, Laboratory of Physics,\\ 
P.O.Box 1100, FIN-02015 HUT, Finland }

\begin{abstract}
The slow motion of a crack line is studied via an experiment in which
sheets of paper are split into two halves in a ``peel-in-nip''
(PIN) geometry under a constant load,  in creep. The velocity-force
relation is exponential.
The dynamics of the fracture line exhibits intermittency, or avalanches,
which are studied using acoustic emission. The
energy statistics is a power-law, with the exponent
$\beta \sim 1.8 \pm 0.1$. Both the waiting times between subsequent
events and the displacement of the fracture line imply
complicated stick-slip dynamics. We discuss the 
correspondence to tensile PIN tests and other similar experiments on 
in-plane fracture and the theory of creep for elastic
manifolds.
\end{abstract}
 
\maketitle
\date{\today}

The deformation and fracture of materials is a fascinating
topic as one can explore the physics even without sophisticated
experiments \cite{advphys}. A piece of paper suffices to give ample evidence
for the presence of phenomena that need a statistical description.
One can tear a sheet or crumble it, to observe
that the response is ``intermittent'' and not simply ``smooth''.
\cite{sethrev,Kertesz,oma,santucci}.

The physics of fracture in a material as paper is governed
by two basic ingredients. The structure  and the microscopic
material properties are inhomogeneous, while the stress fields
follow the laws of elasticity. Typical statistical signatures
that show features that emerge from their interaction
are the acoustic emission during a deformation
test and the post-failure properties of fracture surfaces. These
are commonly found to be described by power-laws, as regards the
energies of acoustic events (``earthquakes''), the intervals between
subsequent events (``Omori's law'') \cite{ciliberto,oma}, 
and the same is true
for the geometry of cracks where ample evidence points towards
a self-affine fractal scaling of the surface fluctuations around
the mean, which is found in a variety of cases (loading conditions,
materials and so forth) \cite{Mandel,Bourev,ponson}. 

The most simple case of fracture - which we also will study here - is the
passage of a {\em crack line} through a sample, when its
movement is constrained on a plane. In this
case, the mathematical description is given by
a crack position $h(x,t)$, where $x$ is along the average
projection of the crack.
The average motion is described by $\bar{h} = vt$
with $v$ the crack velocity. In the situation at hand one
has three ingredients: a disordered environment which
poses obstacles to the line motion, a driving force $K_{ext}$
(stress intensity factor in fracture mechanics language),
and the self-coupling of the fluctuations in $h$ through
a long-range elastic kernel \cite{Fisher}, which is expected
to scale as $1/x$. 

To describe the line's physics one needs the language
 of statistical mechanics. One finds a phase diagram
for $v(K_{ext})$: an immobile crack begins to move
at a critical value $K_c$ of $K_{eff}$ such that for
$K > K_c$ $v>0$. This transition between a ``mobile line''
and a pinned line is commonly called in non-equilibrium statistical
mechanics a ``depinning transition''. Close to the 
critical point $K_c$ the line geometry is a self-affine fractal
with a roughness exponent $\zeta$. This is an example of
a wide class of similar problems, which range from fire
fronts in combustion to flow fronts in porous media to
domain walls in magnets. The planar crack \cite{Fisher2}
 or the contact line
on a substrate \cite{cl} problem has
been studied both theoretically via renormalization
group calculations and numerical simulations, and via
experiments. There is a discrepancy between these, in
that the roughness exponents are $\zeta_{theory} \sim 0.39$
and $\zeta_{expt} \sim 0.6$ \cite{maloy,krauth}. In three-dimensional
experiments the in-plane roughness has recently shown signatures of 
$\zeta_{expt} \sim \zeta_{theory}$ \cite{bonamy}.
Imaging experiments
have shown the existence of {\em avalanches}, of localized
intermittent advances of the crack front with an avalanche
size $s$ distribution $P(s) \sim s^{-1.6 \dots -1.7}$ \cite{maloy}

\begin{figure}
 \includegraphics[width=7cm
,angle=-90]{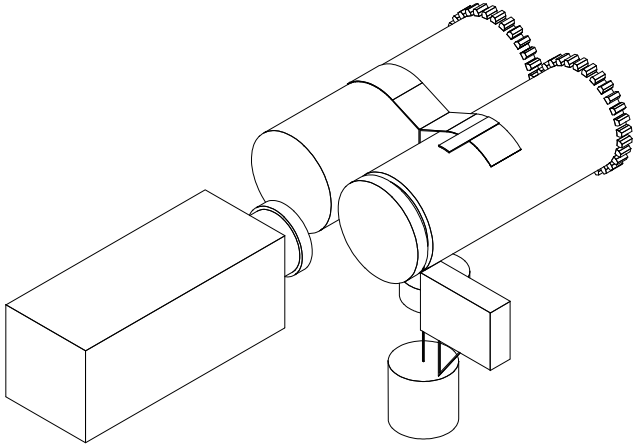}
 \includegraphics[width=7cm
,angle=-90]{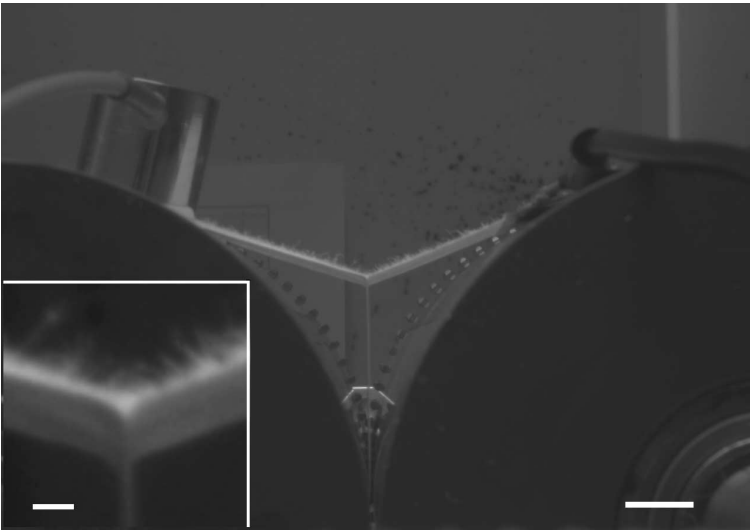}
\caption{Schematic view of the experimental
setup (rolls, paper, AE sensor, camera, weight), 
and two closer views of the geometry.}
\label{fig1}
\end{figure}

Our work considers the dynamics of such a crack front
during {\em creep}, which is done by peeling paper sheets in
a geometry illustrated in Fig.~\ref{fig1} (see also \cite{anantha}).
 The creep
of elastic lines (or manifolds or domain walls) is important
since when $K_{eff} \leq K_c$ thermal fluctuations take over
for $T>0$ \cite{creep,chauve,giatheo}.
The fluctuations nucleate ``avalanches'' similarly to what happens in
zero-temperature depinning in the vicinity of $K_c$. The avalanches
induce a finite velocity $v_{creep}$. The advancement of a crack front
might be fundamentally different from say a domain wall or a contact line, 
which can fluctuate back and forth around a metastable state: it is 
easy to see that the crack may in many cases only grow forward.

The form of $v_{creep} (K_{eff},T)$ depends on the ``energy landscape''
since the current understanding is that creep takes place via
nucleation events over energy barriers \cite{creep}. These barriers are
related to the roughness exponent $\zeta$ and to its origins.
The important physics is summarized with the creep
formula
\begin{equation}
v_{creep} \sim \exp{-(C/m^\mu)}
\label{vcreep}
\end{equation}
which states that the creep velocity is related to the 
driving force, $m$ (as for mass, see Fig. \ref{fig1} again),.
There is  a {\em creep exponent}
$\mu$, which depends on the interactions and dimension of the moving
object (a line). One expects the scaling
\begin{equation}
\mu = \theta/\nu = \frac{\alpha - 2 + 2\zeta}{\alpha-\zeta}
\label{expo}
\end{equation}
in $d$ dimensions.
($d=1$ for a line, $d=2$ for a domain wall in say a bulk magnet).
$\theta$, $\nu$, and $\zeta$ denote the energy fluctuation,
correlation length, and roughness exponents relevant for
the problem, and exponent relations are usually postulated
between these three. They exponents all depend  on $\alpha$, the
$k$-space decay exponent of the elastic kernel. For long range
elasticity, one would assume $\alpha=1$ whereas for the so-called local case 
$\alpha=2$ is expected. Expressed as above, the question
of the value of $\mu$
boils down to the values of $\zeta$ and $\alpha$ - what is the
effective roughness at hand, and what is the elasticity of the
line like? 

The fundamental formula of Eq.~(\ref{expo}) has been confirmed
in the particular case of 1+1-dimensional domain walls \cite{lemerle}, and 
further experimental studies exist \cite{crex}. Our main results on fracture
line creep concern the velocity-force-relation $v(m)$, and on the
picture of the dynamics that ensues, in particular from Acoustic
Emission (AE) data of the avalanches and their dynamics. We find 
an exponential $v(m)$ and discuss its interpretations. The
dynamics shows signatures of intriguing but weak correlations,
and we discuss the similarities and differences to the tensile
case.

In Figure \ref{fig1} we show the apparatus \cite{epl}.
The failure line can be located along ridge, in
center of the the Y-shaped construction formed by the 
unpeeled part of the sheet (below) and the two parts
separated by the advancing line. Diagnostics consist
of an Omron Z4D-F04 laser distance sensor for 
the displacement, and a standard plate-like piezoelectric sensor \cite{epl}. 
It is attached to the
setup inside one of the rolls visible in Fig.~\ref{fig1}, 
and the signal is filtered
and amplified using standard techniques. The data 
acquisition card has four channels at 312.5 kHz
per channel. We finally threshold the AE data. The
displacement data is as expected highly correlated with
the corresponding AE, but the latter turns out to be much
less noise-free and thus convenient to study.
For paper, we use perfectly standard copy paper,
with an areal mass or basis weight of 80 g/m$^2$.
Industrial paper has two principal directions, called
the ``Cross'' and ``Machine'' Directions (MD/CD) \cite{an}.
The deformation characteristics are much more ductile
in CD than in MD, but the fracture stress is higher
in MD. We tested a number of samples for both directions,
with strips of width 30 $mm$. The weight used for the
creep ranges from 380 g (CD) to 533 g (MD), and the resulting
data has upto to tens of thousands of AE events (avalanches)
per test (Fig.~\ref{fig2}). It is unforunately not possible
to infer the critical depinning $m_c > m_{used}$.
The mechanical (and creep) properties of paper depend
on the temperature and humidity. In our setup
both remain at constant levels during an experiment,
and the typical pair of values for the environment is 
49 rH and 27 $^o$C (though we also have data for other 
combinations).

Our main result is given in Figure \ref{fig3}, where
we show the $v(m)$ vs. $1/m$ characteristics.
There are four main datasets depicted. These differ
slightly in the typical temperature and humidity
(for set 2 44 \% rH instead of about 49 \% rH for the others).
These all imply an {\em exponential}
behavior, and by fits to the data sets
we can infer a creep exponent $\nu = 1.0 \pm 0.1$.
This lends itself to two different interpretations: either
the creep is in the 1d random field (RF) domain wall
universality class, since with $\alpha=2$ $\nu=1$ implies $\zeta=1$,
the roughness exponent of the RF universality class at
very small external fields/forces. This assumes
{\em local elasticity} being dominant. 
If we would then
assume that $\zeta$ takes the value $\zeta \sim 0.6\dots0.7$,
of the avalanche regime, this would imply screened interactions
with $\alpha \sim 1.4$.
The thirdr possibility is to use
non-local line elasticity, with $\alpha=1$. Then Eq.~\ref{expo}
indicates that $\zeta \sim 1/3$. This is exactly the {\em equilibrium}
exponent of lines with long-range elasticity \cite{led}.
The velocity in the case of ``paper'' is influenced by the
humidity:  this is clear in our case. Meanwhile, note that the temperature
variation is insignificant here.

The avalanche behavior is illustrated by Fig. \ref{fig3}a, by
the way of the AE event energy distribution $P(E)$. We present
two kinds of avalanche data: integrated over 1 $ms$ windows ($E_i$ 
where $i$ is an integer time index for CD) and extracted
from single avalanches. The former obviously sums over all the avalanches
during the 1 $ms$ duration (if more than one are present). The data
agrees 
rather well regardless of the mass $m$ and the ductility (CD, MD)
with the scaling $P(E)\sim E^{-\beta}$ with $\beta = 1.7\pm 0.1$. 
This $\beta$ is close to the one observed in the corresponding tensile
experiment (depicted in the Fig. \ref{fig3}a) \cite{epl}. 
The simplest interpretation would
be that once an avalanche is created through a thermal fluctuation, it
follows a deterministic course. This is similar to the predictions
of recent theories, which indicate the presence of a small nucleation
scale, and that the avalanches should be as at the depinning critical
point \cite{chauve,giatheo,gia2}.

We have studied the temporal statistics and correlations
using the AE timeseries, in particular the windowed one,
and the direct displacement vs. time -signal (note that the AE
has better accuracy in the time domain). The waiting
times $\tau$, the silent intervals between either two avalanches
or two windows with $E_i>0$ are shown in Fig. \ref{fig3}b. It is
interesting to note how the distributions 
$P(\tau)\sim \tau^\alpha$ change
with the applied force. For large driving forces
it appears that there are (perhaps) two power-law like
regimes: one with an exponent $\alpha = 1.3$, and more clearly
one for large $\tau$ with $\alpha \sim 2$ which is also found
for $m$ small. The general form for $m$ large resembles also
that of similar tensile tests where $\alpha \sim 1$ is
found. In the tensile case, there is a typical 
time/lengthscale (arising from paper structure)
visible in tensile crackline tests, which might here
also be related to the change in the slope of $P(\tau)$.

The dynamics of the line exhibits stick-slip: 
the integrated velocity depends on the
window length under consideration. This is already evident
from the $P(E)$ and extends to longer timescales than what
$P(E)$ implies via durations of events. 
A ``stick-slip'' exponent can be extracted also from
the displacement data as well as integrated from the
AE data. It appears that a relation $P(\Delta h) \sim \Delta h^{-1.7}$
arises. This implies the presence of correlations, which we next
study by the energy release rate $T$: the duration of the active
time $T$ as measured as subsequent windows where $E_i>0$. It can be seen
as illustrated in Figure \ref{fig5} that $P(T) \sim T^{-2.7}$, so that
on a millisecond scale (much slower than avalanche durations, but much
faster than the inverse line velocity) there are clear correlations.
These are however {\em weak} in the respect that one can not see
in here or the AE data signatures of ``aftershocks'' or ``precursors'' 
familiar from earthquakes or from the AE activity of dislocation
systems \cite{weiss}. 
One may also see a correlation between the energy released
in the active period vs. its duration, such that $E_{tot} \sim T^{0.25}$
at least for $m$ small. Thus it appears that the temporal
dynamics can be described by a temporal self-affine process. 

To conclude, we have studied the creep dynamics of an elastic line,
or a fracture front in peeling paper sheets. 
The particular features of our case are the 
disorder present in usual paper and the variation in
material properties (ductile/brittle). The main findings are fourfold.
First, we have obtained an exponential velocity-force relation,
which has three interpretations; we prefer the one which assumes
non-local elasticity implying that an equilibrium $\zeta \sim 1/3$
governs the creep.
Second, we observe intriguing similarities and differences in AE
or avalanches to ordinary
tensile (constant, small velocity) experiments. These, third,
indicate the
separation of deterministic, zero-temperature dynamics from the
nucleation - as in creep indeed - of individual events. Fourth, the 
dynamics of the process exhibits correlations that would need
a theoretical explanation. Our results clearly call for more
theoretical effort to understand in-plane fracture fronts, and
their creep properties. They also indicate the need for general
studies of creep fracture as a statistical physics problem.
{\bf Acknowledgments:} 
The Center of Excellence program of the Academy of Finland 
is thanked for financial support as well as KCL Ltd and the funding
of TEKES,
Finnish Funding Agency for Technology and Innovation. R. Wath\'en, J. Lohi, and K. Niskanen contributed both with moral and practical support. L. Ponson
is acknowledged for a crucial reminder.

\begin{figure}
 \includegraphics[width=7cm
,angle=0]{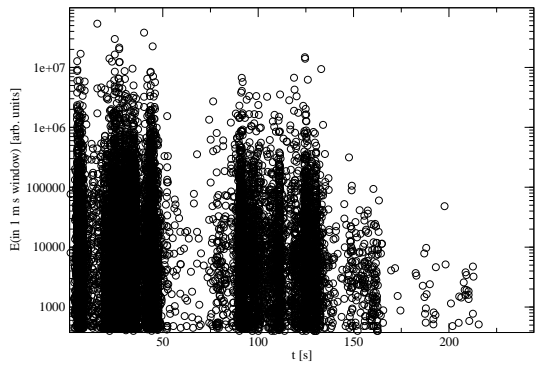}
\caption{An example of the acoustic or stick-slip activity ($E_i>0$)
during
one single creep test. The energy $E_i$ is integrated after
thresholding over 1 $ms$ windows.}
\label{fig2n}
\end{figure}

\begin{figure}
 \includegraphics[width=7cm
,angle=0]{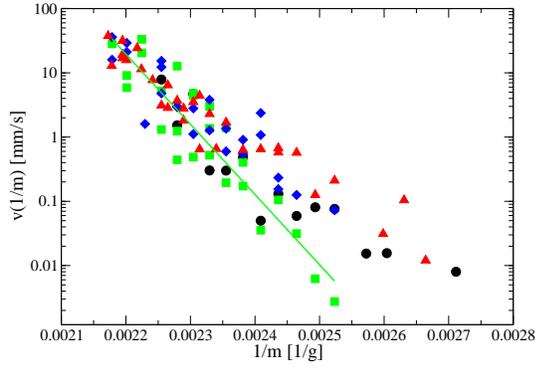}
\caption{The creep velocity vs. the inverse of the
applied force or mass. The line indicates
an exponential decay ($v \sim \exp{a/m}$). The four
sets (circle,rectangle,diamond,triangle) differ
such that the 2nd one was done under lower relative 
humidity, rH \% 43).}
\label{fig2}
\end{figure}

\begin{figure}
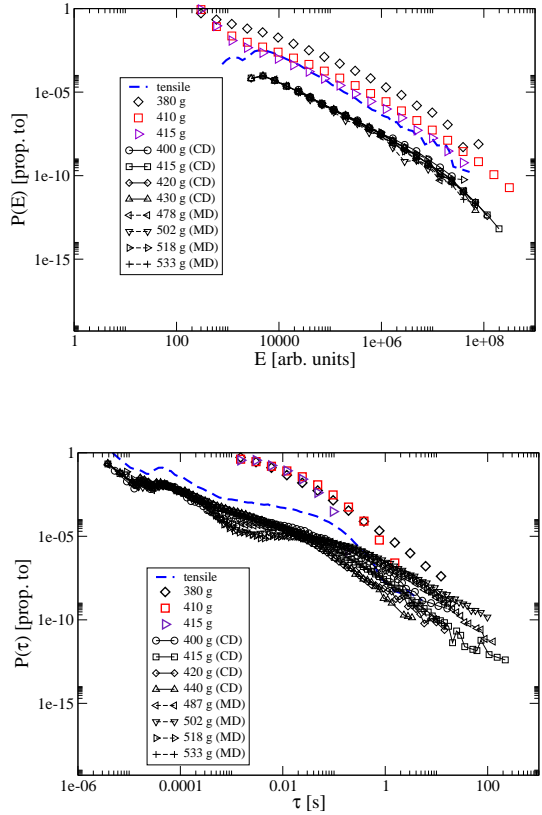

\includegraphics[width=7cm
,angle=0]{./fig4a.eps}
\vspace{10mm}

\includegraphics[width=7cm
,angle=0]{./fig4b.eps}
\caption{a): The probability distributions of 
the acoustic events for various cases. b):
pots of the event interval (waiting time)
distributions, for different massses $m$.
The data sets have been shifted for clarity.}
\label{fig3}
\end{figure}

\begin{figure}
 \includegraphics[width=7cm,angle=-0
,angle=0]{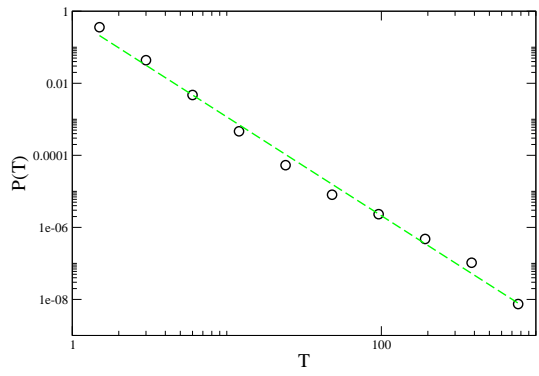}
\caption{The $P(T)$ of durations of active times  $T$ (for definition of $T$
see text) The line has slope -2.7.}
\label{fig5}
\end{figure}

\end{document}